\documentclass[showpacs,preprintnumbers,amsmath,amssymb,      
twocolumn, 
1tightenlines,prl
]{revtex4}

\usepackage{graphicx}
\usepackage{bm}
\usepackage{amsmath}
\usepackage{dcolumn}
\usepackage{epsfig}

\begin{document}

\bibliographystyle{apsrev}

\title { Exponential enhancement of nuclear reactions in condensed
matter environment }

    \author{M.Yu.Kuchiev}
    \email[Email:]{kmy@newt.phys.unsw.edu.au}

    \author{B.L.Altshuler}
    \email[Email:]{bla@feynman.princeton.edu}

    \author{V.V.Flambaum}
    \email[Email:]{flambaum@newt.phys.unsw.edu.au}

    \affiliation {$^* \ddag$University of New South Wales, Sydney,
      Australia}
    
    \affiliation{$\dag$Physics Department, Princeton University,
      Princeton, NJ 08544;\\ NEC Research Institute, 4 Independence
      Way, Princeton NJ 08540, USA}
    
    \affiliation{$\ddag$Institute for Advanced Study, Einstein Drive,
      Princeton, NJ 08540, USA}

        \date{\today}

   \begin{abstract}
     
     A mechanism that uses the environment to enhance the probability
     of the nuclear reaction when a beam of accelerated nuclei
     collides with a target nucleus implanted in condensed matter is
     suggested.  The effect considered is exponentially large for low
     collision energies.  For t + p collision the mechanism becomes
     effective when the energy of the projectile tritium is below
     $\sim$ 1 Kev per nucleon. The gain in probability of the nuclear
     reaction is due to a redistribution of energy and momentum of the
     projectile in several ``preliminary'' elastic collisions with the
     target nucleus and the environmental nuclei in such a way that
     the final inelastic projectile-target collision takes place at a
     larger relative velocity, which is accompanied by a decrease
     of the center of mass energy.  The gain of the relative velocity
     exponentially increases the penetration through the Coulomb
     barrier.
\end{abstract}



\maketitle

   It is well known that nuclear reactions at low energies are
   suppressed by the Coulomb repulsion between the nuclei.  Recent
   experimental papers \cite{1,2,Pd,Ichimaru} suggest that the solid
   state environment of the target nucleus can, possibly,
   significantly enhance the cross section of the DD fusion. The
   previously discussed mechanisms of the enhancement \cite{AFKZ,KAF}
   are efficient only when the energy is too low for the fusion to be
   observable in modern experiments. Here we examine the
   nonsymmetrical collisions when the projectile nucleus is heavier
   than the target. We show that in this case the environment produces
   an exponential enhancement of the cross section of a nuclear
   reaction that has a chance to be observed experimentally.

   We consider a nuclear reaction that is due to a collision of a
   beam of the projectile nuclei with the target nucleus that is
   implanted in a condensed matter environment. The projectile
   energy is presumed to be below the Coulomb barrier, where the
   probability of the nuclear reaction is proportional to the Coulomb
   suppression factor
\begin{equation}\label{supp}
P(v)  = \frac{2\pi Z_\mathrm{proj} Z_\mathrm{tar}e^2}{\hbar v}\exp\left( 
-\frac{2\pi Z_\mathrm{proj} Z_\mathrm{tar}e^2}{\hbar v} \right)~,
  \end{equation}
  here $Z_\mathrm{proj}$ and $Z_\mathrm{tar}$ are the charges of the
  projectile and target nuclei and $v$ is their relative velocity.
  For the collision in the vacuum this velocity equals the initial
  velocity of the projectile $V$, $v=V$. However, the condensed matter
  environment alters the situation because the velocity of the
  collision can change due to redistribution of the momentum and
  energy of the projectile in collisions with the environment nuclei
  and the target nucleus.  We show below that a chain of
  (quasi)elastic collisions with the environment nuclei can, in fact,
  increase the collision velocity. This, according to Eq.(\ref{supp}),
  gives an exponential gain in the probability of the nuclear
  reaction.  However, the probability for the projectile and the
  target to remain on the collision course after collisions with the
  environment nuclei is low.  The more elastic collisions take place,
  the lower it is.  Nevertheless, for sufficiently low collision
  energies the exponential gain due to the increase of the velocity
  inevitably prevails.
  
  The interest in studying the role of the condensed matter
  environment in nuclear reactions is inspired by a few mentioned
  publications, see \cite{1,2,Pd,Ichimaru} and references therein,
  that claim an increase of the DD fusion cross-section in solids.
  Several possible mechanisms that increase the collision velocity in
  the environment were considered previously
  Refs.\cite{KAF,AFKZ,fiorentini_etal_03}. One of them is based on the
  motion of the target nuclei that is due to the phonon vibrations
  (ground-state or thermal) of atoms in solids
  \cite{AFKZ,fiorentini_etal_03}, or nuclear motion inside the atom
  \cite{fiorentini_etal_03}. Another one involves a sequence of three
  elastic collisions \cite{KAF}. This sophisticated chain of events,
  called a ``carambole'' collision in \cite{KAF}, produces a gain for
  the nuclear reaction, but this happens for very low energies of the
  projectile D (below 0.5 Kev) which were not tested in the mentioned
  experiments. In the present paper we examine collisions of the
  projectile nucleus that is heavier than the target nucleus.  In this
  case there exists a {\em rescattering} mechanism that increases the
  collision velocity and relies on only two preliminary elastic
  collisions. This more simple chain of events proves to be much more
  effective than the carambole mechanism of Ref.\cite{KAF}.
  
  There is a long standing discrepancy between the experimental data
  on astrophysical fusion reactions at low energies
  Refs.\cite{engstler_88,engstler_92,angulo_93,preti_94,greife_95} and
  calculations of fusion probabilities. The latter include the effects
  of electron screening, vacuum polarization, relativity,
  bremsstrahlung and atomic polarization, see Refs.
  \cite{assenbaum_etal_88,deglinnocenti_etal_94,balantekin_etal_97}
  and references therein.  The theoretical data slightly, but
  systematically underestimate the probability of the fusion. It is
  unlikely that the rescattering mechanism can be held responsible for
  this discrepancy because the astrophysical data are related to
  nuclear reactions at a given temperature, whereas here we discuss
  the beam-target experiments, where the beam is not thermalized.
  Therefore the rescattering can be relevant only to a non-thermalized
  media; for example to events in the vicinity of a supernova in
  astrophysics, or to the laser induced thermonuclear synthesis in
  laboratory experiments.

\begin{figure}
\centering
 \includegraphics[height=3.4 cm,
keepaspectratio = true, 
]{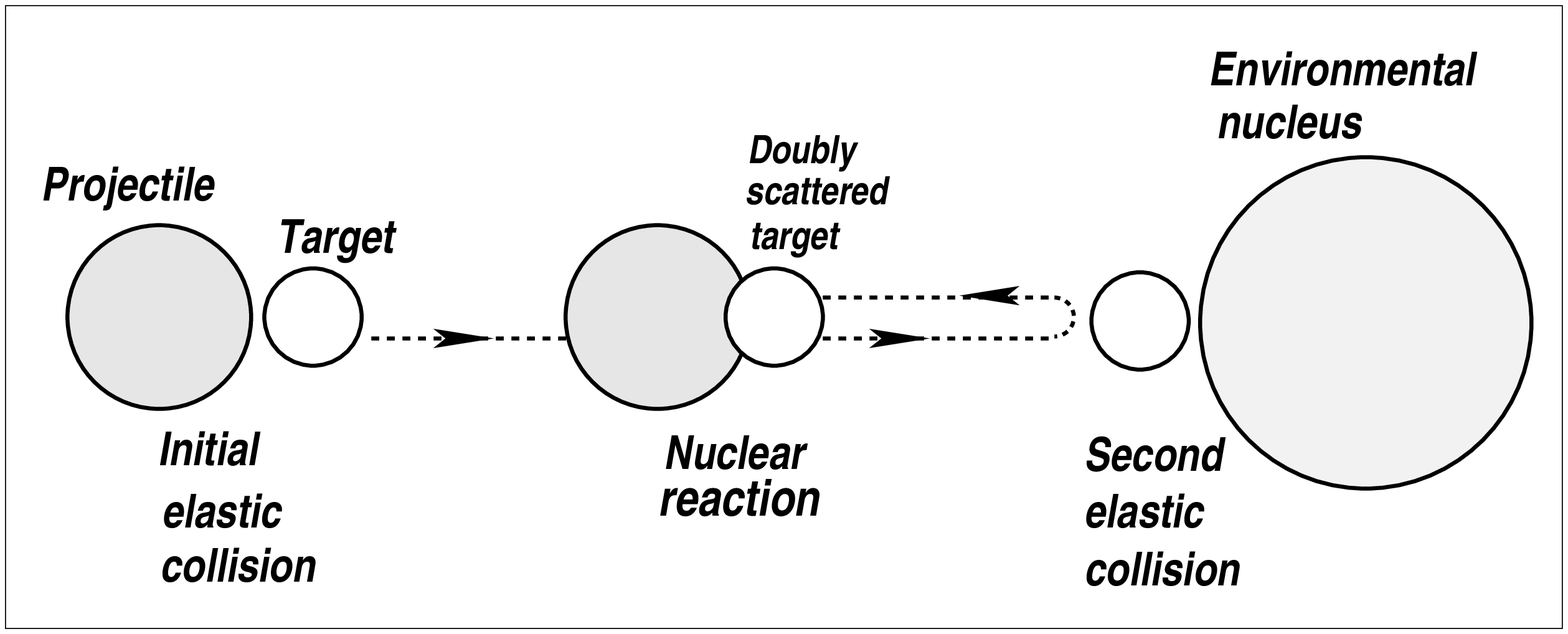}
\caption{ \label{one} Collision of the projectile nucleus with the
  target nucleus in the solid state environment. The rescattering
  mechanism involves two ``preliminary'' elastic collisions.  Firstly
  the elastic collision between the projectile and the target nuclei
  takes place. Then, the recoiled target nucleus collides with the
  heavy nucleus of the environment. After that the final collision of
  the projectile and the (doubly recoiled) target results in the
  nuclear reaction. This chain of events increases the collision
  velocity making the nuclear reaction more probable.  }
\end{figure}

  Let the projectile and the target nuclei have masses $M$ and $m$
  respectively, and $M>m$.  The masses of the environment nuclei
  will be considered large $M_\mathrm{env} \gg m$. Consider the
  following sequence of events shown in Fig. \ref{one}.
     \noindent 
     The projectile with the original velocity ${\bf V}$ first
     collides elastically with the target nucleus initiating its
     motion in the direction of ${\bf V}$. The final velocities $V'$
     and $v'$ of the projectile and the target, which are 
     parallel to ${\bf V}$, are given by
\begin{eqnarray}\label{Vv}
V' = \frac{1-m/M}{1+m/M} \,V~, \quad v' = \frac{2V}{1+m/M}~.
\end{eqnarray}
   Suppose now that there exists an atom of environment located in the
   path of the recoiled target nucleus, as is shown in
   Fig. \ref{one}. Then there is an opportunity for the target nucleus
   to be scattered backward due to its collision with this heavy
   nucleus. The velocity of the doubly scattered target $v'' = -v' = -
   2V/(1+m/M)$ becomes opposite to the projectile velocity.  After
   these two preliminary collisions the projectile and the target
   nuclei find themselves on the collision course for the second
   time. Let us presume that their second encounter results in the
   nuclear reaction. Note that the relative velocity $w$
   of the target and the projectile in this collision is
\begin{equation}\label{v12}
w = V'-v'' =\frac{3-m/M}{1+m/M}\,V~,
   \end{equation}
   which is larger than the initial collision velocity $V$, $w>V$.
   For a heavy projectile $w \simeq 3 V$. Thus the two preliminary
   elastic collisions produce a substantial increase of the relative
   velocity that results in the exponential increase of the
   probability of the nuclear reaction Eq.(\ref{supp}).
   
   However, there is a damping factor that arises due to a necessity
   for the projectile and the target to remain on the path that leads
   to their final inelastic collisions.  In order to calculate this
   damping factor let us introduce the cross sections for the elastic
   collisions: $ d\sigma_\mathrm{proj,tar}/d\Omega$ will be the
   differential cross section that governs the first elastic collision
   between the projectile and target ($d\Omega$ is the solid angle of
   the recoiled target; we need this cross section for the situation
   when the direction of the velocity of the recoiled target coincides
   with the velocity of the incoming projectile.  In the center of
   mass frame (cmf) this corresponds to the backward scattering). The
   flux $J$ of the projectile and target during their final inelastic
   collision equals
\begin{equation}\label{dump}
J=\frac{1}{b^4}
    \frac{d\sigma_\mathrm{proj,tar}}{d\Omega}\,
\frac{d\sigma_\mathrm{tar,env}}{d\Omega}~,
   \end{equation}
   where $b$ is a path of the target nucleus between the environmental
   nucleus and the point of the final collision with the projectile,
   and $d\sigma_\mathrm{tar,env}/d\Omega$ is the differential cross
   section for the backward scattering of the light target on the
   heavy environment nucleus. From simple kinematics it follows that
\begin{equation}\label{b}
b = \frac{v'-V'}{v'+V'}\,a = \frac{1+m/M}{3-m/M}\,a~.
   \end{equation}
   where $a$ is the distance between the initial location of the
   target nucleus and the environmental nucleus.  For a heavy
   projectile $b \simeq a/3 $.  Eq.(\ref{dump}) can be explained
   without calculations.  For the energy range considered the nuclear
   wavelengths are much smaller than typical distances between atoms
   in condensed matter. This implies that the elastic nuclear
   collisions happen at separations that are much smaller than typical
   atomic separations. In other words, the scattering amplitudes for
   the two preliminary elastic collisions are much smaller than
   separations between atoms.  This allows one to approximate the wave
   functions that govern the two elastic collisions by their
   asymptotes that have the conventional form $\psi_\mathrm{elast}
   \simeq (f/r)\exp (ikr)$, where $f$ is the elastic scattering
   amplitude in the cmf.  Within this approximation one can factorize
   the amplitude of the complicated process into a product of elastic
   scattering amplitudes. Correspondingly, the probability is
   presented as a product of the cross sections in Eq.(\ref{dump}).
   Alternatively, one can validate Eq.(\ref{dump}) on the purely
   classical grounds. The first cross section $
   d\sigma_\mathrm{proj,tar}/d\Omega$ specifies the initial elastic
   collision, while the quantities $
   (d\sigma_\mathrm{proj,tar}/d\Omega)/b^2$ and $
   (\sigma_\mathrm{nuc})/b^2$, where $\sigma_\mathrm{nuc}$ is the
   nuclear cross section (that is not presented explicitly in
   Eq.(\ref{dump}), but will be taken into account later on, see
   Eq.(\ref{F}) ), can be considered as two spherical angles, i. e.
   the two probabilities that define the necessary kinematic
   conditions that allow the final inelastic collision to take place.
    
   It follows from the above discussion that the ratio $F$ of the
   probability of the nuclear reaction in the environment due to the
   rescattering mechanism to the probability of the nuclear reaction
   in the vacuum equals
\begin{eqnarray}\label{F}
F & = & J\,\frac{ P(w) }{ P(V) }  = \frac{1}{b^4}
\frac{d\sigma_\mathrm{proj,tar}}{d\Omega}\,
\frac{d\sigma_\mathrm{tar,env}}{d\Omega} 
\\ \nonumber
&\times&
\frac{V}{ w } \exp\left[
\frac { 2\pi Z_\mathrm{proj} Z_\mathrm{tar}e^2 }{\hbar} \left(
\frac{1}{V}-\frac{1}{ w }\right) \right]~.
   \end{eqnarray}
   Here $Z_\mathrm{proj}$ and $Z_\mathrm{tar}$ are the charges of the
   projectile and the target nuclei, $V$ is the velocity of the
   projectile, $w$ is the velocity of the final projectile-target
   collision Eq.(\ref{v12}) (which is preceded by the two preliminary
   elastic collisions). The Coulomb factors $P(V)$, $P(w)$ arise from
   the nuclear cross sections for the collisions with velocities $V$
   and $w$ respectively. We presume here that the velocity-dependence
   of the nuclear cross section is due entirely to the Coulomb factor,
   which is usually a very good approximation for low-energy nuclear
   reactions.  Eq.(\ref{F}) shows that the mechanism discussed
   provides an exponential enhancement of the nuclear reaction, which
   is moderated by a power-type damping $J$-factor related to the two
   elastic collisions. For sufficiently low projectile energies the
   enhancement always prevails.
   
   In order to evaluate the factor $F$ in Eq.(\ref{F}) one needs to
   find the differential cross sections.  They can be calculated in
   the classical approximation (because the wavelengths of all
   colliding nuclei are small).  In examples discussed below we
   consider the proton as a target. The potential that describes the
   interaction of the proton with some other nucleus should include
   the nuclear Coulomb repulsion that is partly compensated by the
   electron screening.  For light nuclei the screening is
   insignificant since for the energy range considered the scattering
   takes place due to those events that happen at very small
   separations between nuclei.  For heavier nuclei the screening is
   more important. Having this in mind, we consider a model in which
   the internuclear potential is approximated by the interaction of the
   bare proton with the Thomas-Fermi potential of the heavier nucleus.
   Calculating the relative trajectory of the colliding nuclei in this
   potential one finds the elastic cross section \cite{LL1}
\begin{equation}\label{ab}
    \frac{ d\sigma_\mathrm{el} }{ d\Omega } = 
\frac{ \rho(\chi) }{ \sin \chi } \,
\left| \frac{ d\rho }{ d \chi } \right|~,
   \end{equation}
   where $\rho$ is an impact parameter, $\chi$ is the scattering
   angle, $\chi = 180^0$ for our case, both for the projectile-target
   collision and collision of the recoiled target with the
   environmental nucleus.
   
\begin{figure}[h]
\centering
\includegraphics[
height=5.2cm,
keepaspectratio = true, 
angle =0]{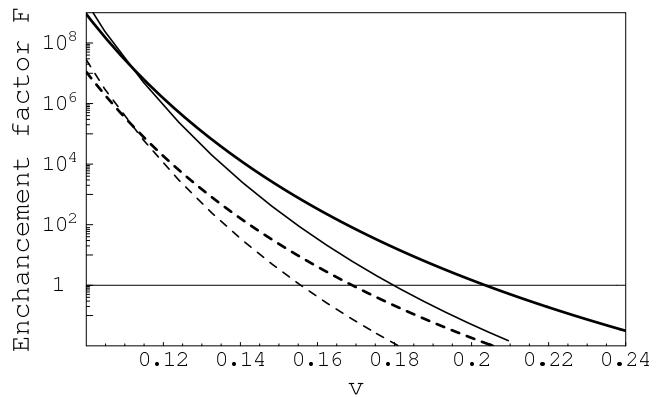}
\caption{ \label{two} 
  Collision of the projectile nucleus with the target proton implanted
  in condensed matter. The enhancement factor $F$ defined in
  Eq.(\ref{F}) is shown versus the velocity per the projectile charge
  (relative units v$ = V/(Z_\mathrm{proj} e^2/\hbar) $ are used).
  Thick lines - tritium as a projectile; solid line $b=1$, dotted line
  $b=3$ in Bohr radiuses, where $b$ is the distance that separates the
  final nuclear event from the environmental nucleus.  Thin lines -
  $^7$Li as a projectile; solid line $b=1$, dotted line $b=3$.}
\end{figure}

   Consider two numerical examples. The first one is the collision of
   the projectile tritium with the proton as a target, with the
   reaction t + p $ \rightarrow $ $^3$He + n, or t + p $\rightarrow$
   $^4$He + $\gamma$.  The rescattering enhances the collision
   velocity by a factor of two, $w=2V$.  Another one uses the $^7$Li
   as the projectile and the proton as a target that leads to the
   reaction $^7$Li + p $\rightarrow \alpha+\alpha$ with the collision
   velocity enhanced due to the rescattering by a factor of 2.5,
   $w=2.5V$. In both cases we assume that the environmental nucleus is
   Pd. (This assumption is not crucial since the elastic cross section
   smoothly depends on the atomic charge of the heavy nucleus.)  We
   estimate the magnitude of the rescattering effect for two values of
   the parameter $b$, taking $b=1$ and $b=3$ in Bohr radiuses (au).
   Fig.  \ref{two} shows the results of calculations of the factor $F$
   that describes the effectiveness of the rescattering mechanism
   comparing it with the reaction in the vacuum. It is shown versus
   the factor $V_\mathrm{proj}/Z_\mathrm{proj}$ that, according to
   Eq.(\ref{supp}), is a natural measure for the probability of the
   reaction.  
     \noindent 
     Fig. \ref{two} shows that, indeed, the rescattering mechanism
     becomes very efficient for sufficiently low projectile energies.
     The rescattering is effective for the t + p collision when
     $V_\mathrm{proj}/Z_\mathrm{proj} \le 0.17-0.2$ au, which
     corresponds to the projectile energy $\varepsilon \le 0.7-1$ Kev
     per nucleon. The energy range down to $\varepsilon \simeq 1$ Kev
     per nucleon was probed for the DD synthesis in Ref.\cite{Pd}.
     This fact points to an opportunity to test the rescattering
     mechanism experimentally in the t + p case in the near future.
     Fig.\ref{two} demonstrates also that the rescattering for the t +
     p collision is more effective than for the $^7$Li + p case. (This
     happens because for a given ratio of
     $V_\mathrm{proj}/Z_\mathrm{proj}$ the elastic cross sections in
     the t + p case are larger than in the $^7$Li + p case.)
    
     Among several factors that were left outside the scope of our
     analysis, probably the most significant one is related to the
     dependence of the results on the geometrical structure that
     describes the actual location of atoms in the condensed matter.
     For the rescattering mechanism to be effective the projectile,
     the target and the environment nucleus need to be aligned.  If
     this condition is slightly violated, then the rescattering
     remains possible, but the relative velocity of the final nuclear
     event becomes smaller.  The stronger the deviation from the
     aligned geometry, the smaller is the collision velocity and less
     effective is the rescattering mechanism. Conscious of this fact,
     we specifically presented data for a sufficiently large parameter
     $b$, $b=3$ au.  For the t + p reaction this corresponds to a
     separation between the proton and the environmental atom in the
     condensed matter $a=6$ au. One can hope that for this large
     separation, deviations from the aligned configuration can be made
     insignificant by taking a suitable condensed matter and selecting
     direction of the projectile beam (though this point should be
     verified more accurately in the future for a particular
     condensed matter environment).
    
     We verified above that the nuclear reaction can be boosted by two
     preliminary elastic collisions. Similarly, one can consider the
     more sophisticated scenario when the target nucleus is
     elastically scattered several, $2n$, $n=1,2 \cdots$ times by the
     target nucleus and the nucleus of the environment. In this
     ``game'' the target nucleus plays the role of a ``ball'' that
     bounces back and forth between the projectile and the
     environmental nuclei $n$ times increasing its velocity with each
     bounce. One can find a similarity between this mechanism and the
     Fermi mechanism of acceleration \cite{FERMI}.  However, there
     exist several factors that restrict the number of bounces. During
     this ``game'' the projectile should keep its velocity in the
     initial direction.  To satisfy this condition it must be
     sufficiently heavy, for $n=2$ the projectile must be at least six
     times heavier than the target, for larger $n$ the mass ratio must
     be even greater.  This restriction rules out sophisticated $n>1$
     cases for the t + p collision.  The ratio of the yield of the
     nuclear reaction after $n$ cycles of bouncing to its yield after
     $n-1$ cycles is proportional to $\propto \exp [( 1 / w_{n-1} - 1
     / w_n )S]$, where $S= 2\pi Z_\mathrm{proj} Z_\mathrm{tar}
     e^2/\hbar$ and $w_n$ is the collision velocity between the
     projectile and the target during their nuclear reaction after $n$
     cycles of elastic rescattering.  This estimate shows that the
     effectiveness of the multiple elastic collisions diminishes with
     the increase of the number $n$ of cycles.  For a sufficiently
     large velocity of the recoil target additional elastic collisions
     make this velocity only slightly larger, while the price of
     additional collisions represented by the damping factor $J$
     (which roughly can be estimated as $n$-independent) remains high.
     Thus multiple collisions are effective only if the simplest $n=1$
     case is very effective, i. e $F\gg 1 $. Therefore multiple
     rescattering can give a contribution to the magnitude of the
     effect, but the fact of the exponential enhancement of the
     probability of the nuclear reaction follows from the simplest
     case, when only two preliminary elastic collisions take place.
     
     In summary, the rescattering mechanism considered proves
     effective. Our estimations for the t + p collision show that when
     the energy of the projectile tritium is in the region of $\sim
     0.7-1$ Kev per nucleon then the probability of the nuclear
     reaction induced by this mechanism exceeds the probability of the
     direct event. For lower energies the discussed mechanism provides
     an exponential boost for the reaction.
   
     The authors are grateful to C.A.Bertulani and V.G.Zelevinski for
     discussions.  M.K. is thankful for hospitality of the staff of
     the School of Physics at the Princeton University where this work
     was completed.  V.F. is grateful to the Institute for Advanced
     Study and Monell foundation for hospitality and support. This
     work was supported by the Australian Research Council and the
     Australian Academy of Sciences.




\begin{thebibliography}{99}
    
  \bibitem{1}

    S. E. Jones, E. P. Palmer, J. B. Czirr , D. L. Decker, G. L.
    Jensen, J. M. Thorne, S. F. Taylor, and J. Rafelski, Nature {\bf
      338}, 737 (1989).

\bibitem{2} 
  
  A. Arzhannikov and G. Kezerashvili, Phys. Lett. A {\bf 156}, 514
  (1991).

\bibitem{Pd} H. Yuki, J. Kasagi, A. G. Lipson, T. Ohtsuki, T. Baba, T.
  Noda, B. F. Lyakhov, and N. Asami, Pis'ma Zh. Eksp. Teor. Fiz., {\bf
    68}, 785 [JETP Lett. {\bf 68}, 823] (1998).

\bibitem{Ichimaru}
  
  S. Ichimaru, {\it Rev. Mod. Phys.} {\bf 65}, 255 (1994).


\bibitem{AFKZ}
  
  B. L. Altshuler, V. V. Flambaum , M. Yu. Kuchiev and V. G.
  Zelevinsky,
  Journal of Physics G - Nuclear Particle Physics {\bf 27}, 2345
  (2001).


\bibitem{KAF}
  
  M. Yu. Kuchiev, B. L. Altshuler, and V.V.Flambaum.
  J.Phys G - Nuclear Particle Physics {\bf 28} 47 (2002).


\bibitem{fiorentini_etal_03} 
  
  G. Fiorentini, C. Rolfs, F. L. Villante, and B. Ricci, Phys. Rev. C
  {\bf 67}, 014603 (2003).


\bibitem{engstler_88}
  
  S. Engstler {\it et al}., Phys. Lett. {\bf B} 2, 179 (1988).


\bibitem{engstler_92}
 
  S. Engstler {\it et al}., Z.Phys. {\bf A} 342, 471 (1992).


\bibitem{angulo_93}
  
  C. Angulo {\it et al}., Z.Phys. {\bf A} 345, 231 (1993).


\bibitem{preti_94}
  
  P.Preti {\it et al}., Z.Phys. {\bf A} 350, 171 (1994).


\bibitem{greife_95} 
  
  U. Greife {\it et al}., Z.Phys. {\bf A} 351, 107 (1995).


\bibitem{assenbaum_etal_88} 
  
  H. J. Assenbaum, K.  Langanke, and G. Soff, Phys. Lett. B {\bf 208},
  346 (1988).


\bibitem{deglinnocenti_etal_94}
  
  S. Degl'Innocenti and G. Fiorentini, Astron. Astrophys. {\bf 284},
  300 (1994).

  
\bibitem{balantekin_etal_97} 
  
  A. B. Balantekin, C. A. Bertulani, and M. S. Hussein, Nucl. Phys.
  {\bf A, 627} (1997).


\bibitem{LL1}
  
  L. D. Landau and E. M. Lifshitz, {\it Classical mechanics}
  (Pergamon, New York) (1976).

\bibitem{FERMI}
  
  E. Fermi, Phys. Rev. {\bf 75}, 1169 (1949).


\end{thebibliography}
\end{document}